\documentstyle[aps,epsfig,multicol,prb]{revtex}
\begin{document}

\draft

\renewcommand{\narrowtext}{\begin{multicols}{2} \global\columnwidth20.5pc}
\renewcommand{\widetext}{\end{multicols} \global\columnwidth42.5pc}

\multicolsep = 8pt plus 4pt minus 3pt

\def\inseps#1#2{\def\epsfsize##1##2{#2##1} \centerline{\epsfbox{#1}}}

\def\top#1{\vskip #1\begin{picture}(290,80)(80,500)\thinlines \put(
65,500){\line( 1, 0){255}}\put(320,500){\line( 0, 1){
5}}\end{picture}}

\def\bottom#1{\vskip #1\begin{picture}(290,80)(80,500)\thinlines \put(
330,500){\line( 1, 0){255}}\put(330,500){\line( 0, -1){
5}}\end{picture}}

\title{Geometric (Berry) phases and statistical transmutation 
 in the two-dimensional systems of strongly correlated electrons}
\author{Seung-Pyo Hong and Sung-Ho Suck Salk}
\address{Department of Physics, Pohang University of Science and Technology,
 Pohang 790-784, Korea}
\maketitle

\begin{abstract}
Focusing on the hole-doped two-dimensional systems of 
strongly correlated electrons,
we examine geometric phases
acquired by electronic wave functions
as a result of polaron transport around a closed loop.
For this study we apply the Lanczos exact diagonalization method to 
Holstein-Hubbard, Holstein-$tJ$, and Holstein-$tJ_z$ models
in order to reveal various aspects of geometric phases.
We demonstrate that transverse spin fluctuations
are responsible for the generation 
of nontrivial geometric phases.
From the exchange symmetry of polarons
we find that the statistical transmutation of polarons occurs
depending on the strength of electron correlations.
\end{abstract} 


\narrowtext

\section{INTRODUCTION}
\vspace*{-3mm}
Since the discovery of high temperature superconductivity
in copper oxides,
attempts have been made to identify
the nature of charge carriers in these materials.
There are some experimental evidences
that polaronic charge carriers exist
in both the superconducting state\cite{zhkm97}
and the normal state.\cite{be93,edsmsyhua94}
Photoinduced absorption measurements
in ${\rm La_2CuO_4}$ and ${\rm YBa_2Cu_3O_{7-\delta}}$
indicate that self-localized structural distortions are present
around photoexcited charge carriers.\cite{kim}
A recent experiment demonstrated that
Sr-induced local lattice distortions occurs in ${\rm La_{2-x}Sr_xCuO_4}$
in association with holes donated by the Sr atoms.\cite{hshmj97}
In order to investigate the nature of the polaronic charge carriers,
we study geometric phases acquired by the electronic wave functions
after the transport of a polaron around a closed loop.

A quantum object (a subsystem of fast degrees of freedom)
acquires a geometric (Berry) phase
in an environment where a parameter
(a subsystem of slow degrees of freedom) is slowly
transported around a closed circuit in a parameter space.\cite{berry84}
Thus quantum subsystems with two widely separated energy scales
(between slow and fast degrees of freedom) 
exhibit geometric phases.\cite{msw89,nakahara90}
In this paper we investigate the geometric phases
acquired by the electronic wave functions
as a result of polaron transport around a closed loop
in a hole-doped two-dimensional system of strongly correlated electrons,
e.g., the copper oxide plane in high $T_c$ superconductors.
In addition, we investigate the statistical transmutation of polarons
by studying the exchange symmetry of polarons
varying with the strength of electron correlations.

\section{VARIOUS GEOMETRIC (BERRY) PHASES BASED ON MODEL HAMILTONIANS}
\vspace*{-3mm}
\label{sec:model}
We first study the geometric phases
varying with the strength of antiferromagnetic electron correlations 
and of electron-phonon coupling.
For this study we introduce 
the Holstein-Hubbard model Hamiltonian
for the hole-doped two-dimensional systems of 
square lattice,\cite{prz87,zs92}
\addtocounter{equation}{1}
\newcounter{ham}
\setcounter{ham}{\arabic{equation}}
$$
 H = -t \sum_{\langle ij \rangle\sigma}
 (c_{i\sigma}^\dag c_{j\sigma}^{} + \mbox{c.c.})
 + U \sum_i n_{i\uparrow} n_{i\downarrow}
 - g \sum_i u_i n_i
$$
\vspace*{-4mm}
$$
 \hspace*{-35mm}
 + \frac{K}{2} \sum_i (x_i^2 + y_i^2)
 ~~,
 \eqno{(\arabic{ham}{\rm a})}
$$
with the Holstein coordinate $u_i$,
$$
 u_i = \frac{1}{4} 
 (x_{i_x,i_y}^{} - x_{i_x-1,i_y}^{} + y_{i_x,i_y}^{} - y_{i_x,i_y-1}^{})
 ~~.
 \eqno{(\arabic{ham}{\rm b})}
$$
Here $c_{i\sigma}^\dag$ ($c_{i\sigma}^{}$) is the creation (annihilation)
operator of an electron with spin $\sigma$ 
at a copper site $i$ ($i=(i_x,i_y)$),
and $n_{i\sigma}^{}=c_{i\sigma}^\dag c_{i\sigma}^{}$
is the electron number operator.
$x_i$ and $y_i$ denote the displacements of oxygen atoms
in the unit cell of ${\rm CuO_2}$ plane.\cite{rfb93,frmb93}
$t$ represents the electron hopping integral;
$U$, the on-site Coulomb repulsion energy (correlation strength);
$g$, the electron-lattice coupling constant
and $K$, the spring constant.
In our calculations 
the magnitude of the Holstein coordinate $u_i^{}$ 
is normalized to unity, $|u_i^{}|=1$
when a polaron is formed around copper site $i$
as a result of hole doping.
The energy of the polaron is then given by $-g$.

Now for the study of geometric phases varying with 
the Heisenberg coupling constant $J$
and the electron-phonon coupling constant $g$,
Hamiltonians of interest are
the Holstein-$tJ$ model,\cite{rfb93,frmb93}
\begin{eqnarray}
 H &=& -t \sum_{\langle ij \rangle\sigma}
 (\tilde{c}_{i\sigma}^\dag \tilde{c}_{j\sigma}^{} + \mbox{c.c.})
 + J \sum_{\langle ij \rangle} {\bf S}_i \cdot {\bf S}_j
 - g \sum_i u_i n_i
 \nonumber \\
 &&
 + \frac{K}{2} \sum_i (x_i^2 + y_i^2)
 ~~,
\end{eqnarray}
and the Holstein-$tJ_z$ model,
\begin{eqnarray}
 H &=& -t \sum_{\langle ij \rangle\sigma}
 (\tilde{c}_{i\sigma}^\dag \tilde{c}_{j\sigma}^{} + \mbox{c.c.})
 + J_z \sum_{\langle ij \rangle} S_i^z S_j^z
 - g \sum_i u_i n_i
 \nonumber \\
 &&
 + \frac{K}{2} \sum_i (x_i^2 + y_i^2)
 ~~.
\end{eqnarray}
Here $\tilde{c}_{i\sigma}^{} = c_{i\sigma}^{} (1 - n_{i,-\sigma}^{})$
is the electron annihilation operator at site $i$,
which excludes double occupation.
${\bf S}_i = \frac{1}{2} c_{i\alpha}^\dag \bbox{\sigma}_{\alpha\beta}^{}
c_{i\beta}^{}$ is the electron spin operator.
The Holstein-$tJ_z$ Hamiltonian takes into account only the 
spin $z$-component interaction.

The lattice distortions 
are treated in the adiabatic limit.\cite{prz87,frmb93}
The electronic wave functions (the fast degree of freedom)
varying with lattice distortions
are obtained from the application of
Lanczos exact diagonalization method\cite{dagotto94}
to a tilted $\sqrt{10}\times\sqrt{10}$ lattice
with periodic boundary conditions.\cite{frmb93}
The propagation of the local lattice distortion 
(polaron hopping) is described by
\begin{equation}
 u_i^{}(\tau) = u_{{\scriptscriptstyle A},i}^{}
 + (u_{{\scriptscriptstyle B},i}^{} - u_{{\scriptscriptstyle A},i}^{})\tau
\end{equation}
with $\tau$ being the dimensionless time lapse with $0 \le \tau \le 1$
for polaron hopping between adjacent lattice sites.
For instance, the lattice distortion occurs at copper site $A$ at $\tau=0$
and moves to an adjacent copper site $B$ at $\tau=1$.
In this manner, a closed path in the parameter space $\{u_i^{}\}$
can be defined.
We compute the geometric phase factors, $e^{i\gamma_n^{}(C)}_{}$
acquired by the electronic wave function
after polaron transport around a closed path $C$ during time $T$,
by using\cite{ki85}
\begin{equation}
 e^{i\gamma_n^{}(C)}_{} =
 \lim_{N \rightarrow \infty} \prod_{k=1}^N
 \langle n(u(\tau_k^{})) | n(u(\tau_{k-1}^{})) \rangle
 ~~.
 \label{factor}
\end{equation}
Here $|n(u(\tau_k))\rangle$ is the electronic eigenfunction
for a given lattice distortion $u(\tau_k)$ at time $\tau_k$.
$\tau_k$ is a $k$th discretized time lapse 
between the initial time $\tau_0^{}=0$ 
and the final time (period) $\tau_{\scriptscriptstyle N}^{}=T$.
Eq.~(\ref{factor}) is derived under the condition that
$|n(u(\tau_k))\rangle$ is a single-valued complex wave function,
that is, $|n(u(0))\rangle = |n(u(T))\rangle$.\cite{berry84,ki85}
On the other hand,
all the electronic wave functions obtained from
the diagonalization of the model Hamiltonians above
are real-valued.
In such cases, geometric phases appear 
in the form of double-valued real wave 
functions.\cite{berry90,aitchison88,mt79,hl63}
Using the local gauge transformation in parameter space $u$,
the double-valued real wave function, $|n(u)\rangle_r$ 
can be converted into a single-valued complex wave function $|n(u)\rangle_c$,
\begin{equation}
 |n(u)\rangle_c = e^{-i\theta(u)}_{} |n(u)\rangle_r
\end{equation}
with $\theta(u)$, a differentiable function and
$\theta(u(T)) - \theta(u(0)) = \pi$ (see Appendix for details).
By introducing the above expression into Eq.~(\ref{factor}),
one can now obtain the geometric phases of interest.

\section{COMPUTED BERRY PHASES AND STATISTICAL TRANSMUTATION}
\begin{figure}[htb]
 \centering
 \epsfig{file=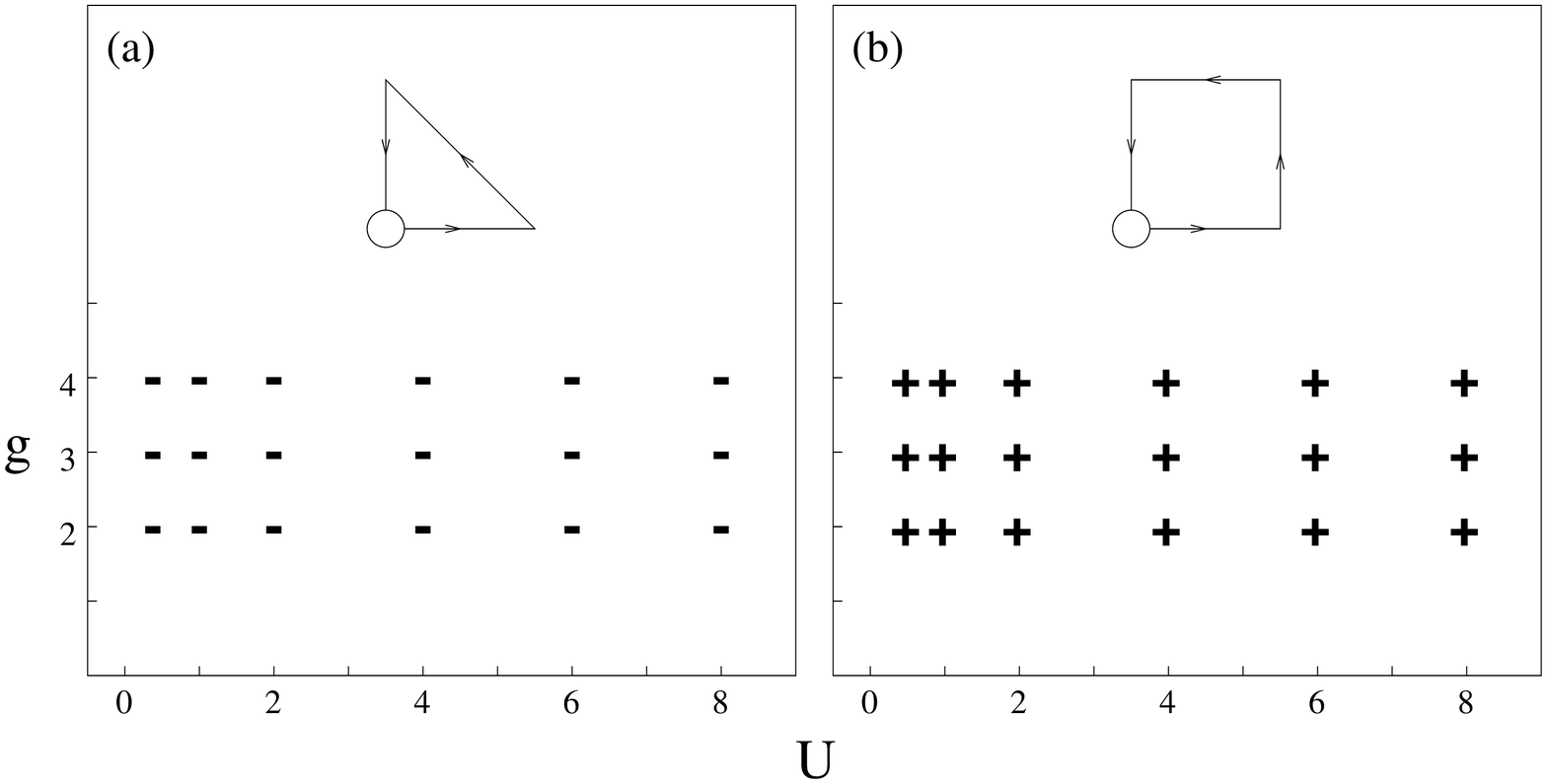, width=7cm}
 \caption{
  Geometric phase factors
  based on the Holstein-Hubbard model
  for (a) a triangular loop and (b) a square loop.
  The horizontal axis denotes the Coulomb correlation strength $U$
  and the vertical axis, the electron-lattice coupling constant $g$.
  The energy unit is $t=1$.}
 \label{phaseU}
\end{figure}

\begin{figure}[htb]
 \centering
 \epsfig{file=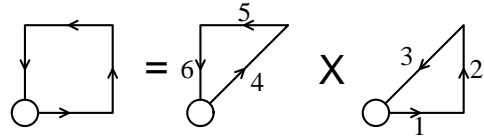, width=7cm}
 \caption{
  The decomposition of a square path into two triangular paths.}
 \label{decompose}
\end{figure}

First, by using the Holstein-Hubbard model
we consider the Berry phase
acquired by the electronic wave function
as a result of polaron hopping around a closed path.
Two types of closed paths,
a triangular path and a square path
are displayed in Figs.~\ref{phaseU}(a)--(b).
For simplicity, for the next nearest neighbor hopping
we chose the identical value of hopping integral 
to the one used for the nearest neighbor hopping.
The electronic wave function is predicted to gain 
a geometric phase by $\pi$
(the geometric phase factor of $-1$)
for polaron hopping around the smallest possible closed path, that is,
the triangular path [Fig.~\ref{phaseU}(a)].
On the other hand,
the electronic wave function
with the polaron transported around the square path [Fig.~\ref{phaseU}(b)]
is predicted to gain a phase angle by $2\pi$,
thus leaving the phase factor of the electronic wave function unchanged.
The geometric phase of $2\pi$ 
can be understood from the decomposition of
the square path into two triangular paths,
as shown in Fig.~\ref{decompose};
the geometric phase factor of $+1$ for the square path results from
the product of two identical geometric phase factors of $-1$
for the decomposed triangular paths.

\begin{figure}[htb]
 \centering
 \epsfig{file=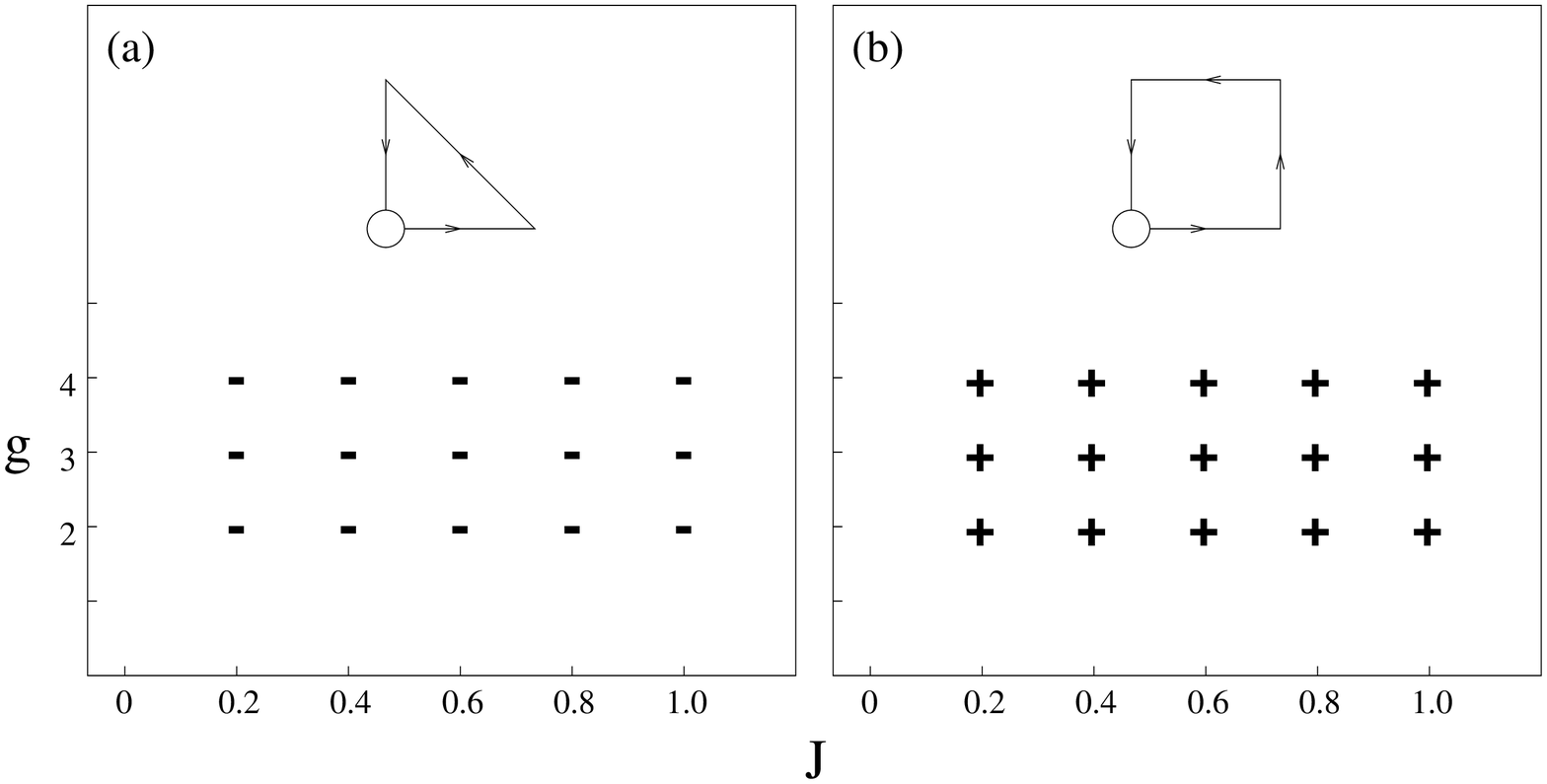, width=7cm}
 \caption{
  Geometric phase factors 
  based on the Holstein-$tJ$ model
  for (a) a triangular loop and (b) a square loop.
  The horizontal axis denotes the Heisenberg exchange interaction $J$
  and the vertical axis, the electron-lattice coupling constant $g$.
  The energy unit is $t=1$.}
 \label{phaseJ}
\end{figure}

\begin{figure}[htb]
 \centering
 \epsfig{file=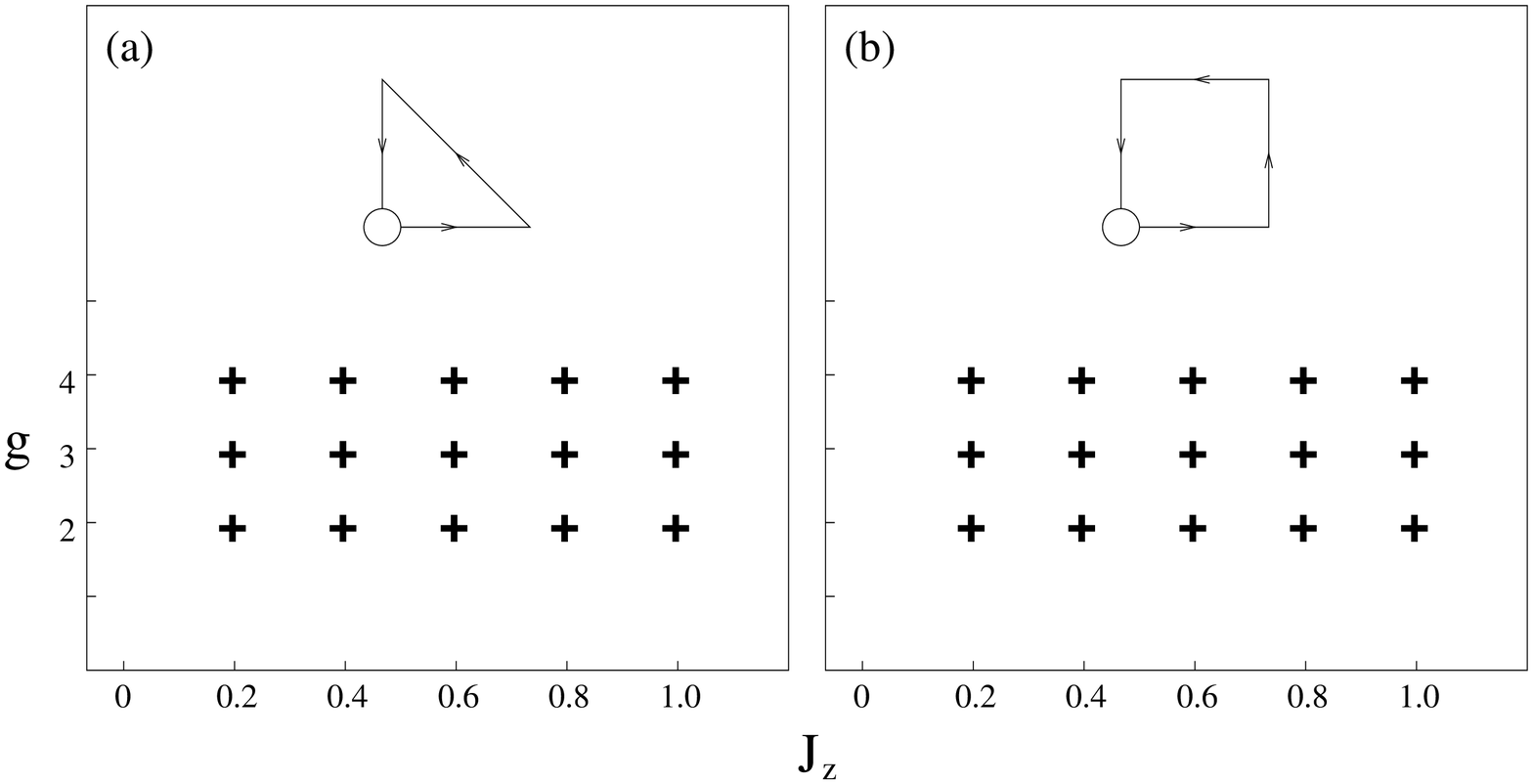, width=7cm}
 \caption{
  Geometric phase factors 
  based on the Holstein-$tJ_z$ model
  for (a) a triangular loop and (b) a square loop.
  The horizontal axis denotes the Ising interaction $J_z$
  and the vertical axis, the electron-lattice coupling constant $g$.
  The energy unit is $t=1$.}
 \label{phaseJz}
\end{figure}

By using both the Holstein-$tJ$ model
and the Holstein-$tJ_z$ model,
we now examine the geometric phases 
which result from the transport of a single polaron.
They are shown in Figs.~\ref{phaseJ}(a)--(b) and Figs.~\ref{phaseJz}(a)--(b).
The predicted geometric phase factor
for the case of the triangular path [Fig.~\ref{phaseJ}(a)]
is in agreement with that of Sch\"{u}ttler {\it et al}.\cite{syz95}
For large $U$ values, say, $U=8t$
both the Holstein-$tJ$ model and the Holstein-Hubbard model yielded
identical geometric phase factors
for both the triangular path and the square path,
as are shown in Figs.~\ref{phaseU}(a)--(b) and Figs.~\ref{phaseJ}(a)--(b).
Such direct comparison between the two model Hamiltonians
is valid only for large $U$ values.
This is because
the $t$-$J$ model is equivalent to the Hubbard model Hamiltonian
corresponding to large $U$ values.
On the other hand,
the Holstein-$tJ_z$ model yields 
a trivial geometric phase factor of $+1$
for the triangular path [Fig.~\ref{phaseJz}(a)],
while the Holstein-$tJ$ model Hamiltonian gives rise to 
the nontrivial geometric phase factor of $-1$.
It is of note that only the $t$-$J$ (but not the $t$-$J_z$) Hamiltonian
allows transverse spin fluctuations.
Thus we discover from comparison of the two Hamiltonians that
the generation of the nontrivial geometric phase factor of $-1$ is caused by
the transverse spin fluctuations
(or spin flip-flop fluctuations) 
which comes from interactions involving the $x$- and $y$-component
(but not the $z$-component) spins, that is,
$J\sum_{\langle ij\rangle} (S_{ix} S_{jx} + S_{iy} S_{jy}) =
 \frac{1}{2} J\sum_{\langle ij\rangle} (S_{i+} S_{j-} + S_{i-} S_{j+})$.
The nontrivial geometric phases
are also predicted by the Holstein-Hubbard model Hamiltonian,
in agreement with the Holstein-$tJ$ model Hamiltonian,
as is shown in Fig.~\ref{phaseU}(a).

\begin{figure}[htb]
 \centering
 \epsfig{file=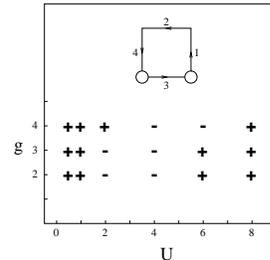, width=3.5cm}
 \caption{
  Geometric phase factors
  based on the Holstein-Hubbard model
  for two-polaron exchange around a square path.
  The numeral label for the path indicates
  the consecutive order of polaron hopping.
  The axes are the same as in Fig.~\ref{phaseU}.}
 \label{phaseU2}
\end{figure}

Now we investigate the exchange symmetry of two polarons
based on the Holstein-Hubbard model.
The numeral label for the square path in Fig.~\ref{phaseU2} indicates
the consecutive order of polaron hopping.
Interestingly enough,
there exist both trivial and nontrivial geometric phase factors.
After the two polarons are exchanged,
the electronic wave function is seen to acquire
a trivial geometric phase factor of $+1$
for the case of weak electron correlations
($U \raisebox{-0.5mm}{$\,\stackrel{\scriptstyle <}{\scriptstyle \sim}\,$} 1$).
This is in sharp contrast with the nontrivial geometric phase factor of $-1$
for the case of intermediate electron correlations
($2 \raisebox{-0.5mm}{$\,\stackrel{\scriptstyle <}{\scriptstyle \sim}\,$}
  U \raisebox{-0.5mm}{$\,\stackrel{\scriptstyle <}{\scriptstyle \sim}\,$} 4$).
A reentrant behavior of the trivial geometric phase factor of $+1$ occurs
for the case of strong antiferromagnetic electron correlations
($U \raisebox{-0.5mm}{$\,\stackrel{\scriptstyle >}{\scriptstyle \sim}\,$} 6$).
It is thus proper to state that
polarons behave as hard core bosons 
in the case of weakly and strongly correlated electrons,
while they act as fermions in the intermediate region of
electron correlations.
Such statistical transmutation of polarons
can be mapped into the Jordan-Wigner transformation\cite{fradkin91}
for the two-dimensional lattice,
\begin{equation}
 a_i^{} a_j^\dag = \delta_{ij}^{} - e^{i\delta}_{} a_j^\dag a_i^{} ~~.
\label{jw}
\end{equation}
We find that the predicted phase angle $\delta$ depends
on $U$, the strength of electron correlation (Coulomb repulsion).
That is, $\delta=\pi$ for small and large $U$ values,
indicating that polarons behave as bosons
and $\delta=0$ for intermediate $U$ values,
indicating that they act as fermions.

\begin{figure}[htb]
 \centering
 \epsfig{file=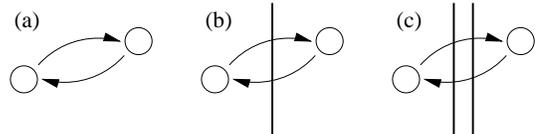, width=7cm}
 \caption{
  Exchange of two polarons. Open circles denote polarons,
  and each vertical solid line represents a fictitious flux line
  of one flux quantum (flux unit).}
 \label{exchange}
\end{figure}

We examine the exchange symmetry of polarons in different aspect.
For the case of weak electron correlations
(or free electron limit,
$U \raisebox{-0.5mm}{$\,\stackrel{\scriptstyle <}{\scriptstyle \sim}\,$} 1$)
the predicted exchange symmetry indicates a bosonic nature of polarons
with the absence of fictitious magnetic flux line.
This is depicted in Fig.~\ref{exchange}(a).
As correlation (repulsive interaction) between electrons increases,
a change of the geometric phase is predicted.
That is,
for the case of intermediate correlations
it can be stated that
the polarons hop around a fictitious flux 
of an elementary flux unit (fluxon),
thereby exhibiting a fermionic nature
(see a solid line in Fig.~\ref{exchange}(b)).
For the case of strong antiferromagnetic electron correlations,
the exchange of the two polarons indicates 
the presence of two fictitious flux quanta,
as is depicted in Fig.~\ref{exchange}(c).
Then the statistical transmutation of boson to a composite boson occurs.
By composite boson we mean that the boson remains as a boson
with the acquired phase of $2\pi$.
In short the number of fictitious flux quanta
is seen to vary with the strength of electron correlation
and allows the statistical transmutation.

\section{CONCLUSION}
\vspace*{-3mm}
By applying the Lanczos exact diagonalization method to 
Holstein-Hubbard, Holstein-$tJ$, and Holstein-$tJ_z$ model Hamiltonians,
we have examined the different aspects of geometric phases
acquired by electronic wave functions
during the transport of a single polaron around a triangular loop.
We find that the nontrivial geometric phase factor of $-1$
is caused by the presence of transverse spin fluctuations.
Further by examining the exchange symmetry of polarons
based on the Holstein-Hubbard model
we discover the statistical transmutation of polarons
which varies with the strength of antiferromagnetic electron correlation.
It is interesting to note 
that polarons behave as bosons 
in the hole-doped two-dimensional system (background) of 
weakly and strongly correlated electrons
and as fermions 
in the background of intermediate correlation strength.
In the future
it remains to examine 
how the statistical transmutation for the exchange of polarons occurs 
as a function of electron correlation
and whether there exists a possibility of 
anyonic exchange phase other than $\delta=0$ and $\delta=\pi$,
that is,  $0 < \delta < \pi$.

\vspace*{-3mm}
\acknowledgments{ \vspace*{-3mm}
S.-H.S.S. acknowledges the financial supports
of Korean Ministry of Education (BSRI-98) 
and of the Center for Molecular Sciences at KAIST.~
We are grateful to Dr. Kwangyl Park for discussions.
}

\vspace*{-3mm}
\appendix
\section*{}
\vspace*{-3mm}
For the case of multi-valued real wave functions
here we prove the Berry phase to be
$\gamma(C)=\theta({\bf R}(T))-\theta({\bf R}(0))$
by introducing a local gauge transformation in the parameter space,
${\bf R}(=u)$.
The Berry phase $\gamma_n(C)$ is a geometric contribution
to the total phase change for the final state of
$|\psi(T)\rangle = \exp[i\gamma_n(C)]
 \exp\left\{-\frac{i}{\hbar}\int_0^T dt E_n({\bf R}(t))\right\}
 |\psi(0)\rangle$
where $T$ is the time period.\cite{berry84}
For a subsystem of fast degrees of freedom 
coupled to a subsystem of slow degrees of freedom
the Berry phase is obtained from 
\begin{equation}
 \gamma_n(C) = i\oint \langle n;{\bf R}|
 \nabla_{\!\scriptscriptstyle R} |n;{\bf R}\rangle \cdot d{\bf R} ~,
\label{bergam}
\end{equation}
~\\[-5mm]
as is well known.\cite{berry84}

The equation of motion for the subsystem of fast degrees of freedom is given by
\begin{equation}
 H({\bf R}(t)) |n;{\bf R}(t)\rangle 
 = E_n({\bf R}(t)) |n;{\bf R}(t)\rangle ~,
\label{eqmotion}
\end{equation}
~\\[-5mm]
where ${\bf R}(t)$ is the slowly varying parameter.
To obtain the nontrivial (non-zero) Berry phase 
in Eq.~(\ref{bergam}), 
the eigenstate $|n;{\bf R}(t)\rangle$
should be complex, non-degenerate, and single-valued.
The gauge potential
\begin{equation}
{\bf A}({\bf R}) = i\langle n;{\bf R}|
 \nabla_{\!\scriptscriptstyle R} |n;{\bf R}\rangle
\label{connection}
\end{equation}
~\\[-5mm]
vanishes for the case of the real eigenstates $|n;{\bf R}\rangle$.

It is obvious that the nontrivial (nonzero) `magnetic field' 
${\bf B}({\bf R})=\nabla\times{\bf A}({\bf R})$
cannot be defined
for the case of real eigenstates $|n;{\bf R}(t)\rangle_r$,
since the gauge potential ${\bf A}({\bf R})$ vanishes.
However, by using the following gauge transformation it is possible
to define the nonvanishing Berry phase:
\begin{equation}
 |n;{\bf R}\rangle_c = e^{-i\theta({\bf R})}
 |n;{\bf R}\rangle_r
 ~~.
\label{gauget}
\end{equation}
~\\[-5mm]
Indeed, this is the local gauge transformation
in the parameter space ${\bf R}$,
which provides the gauge invariance of 
the Schr\"{o}dinger equation, Eq.~(\ref{eqmotion}).
Using Eq.~(\ref{gauget}) in Eq.~(\ref{bergam}) we readily obtain 
the Berry phase $\gamma_n(C)$,
\begin{equation}
\gamma_n(C) = \theta({\bf R}(T)) - \theta({\bf R}(0)) 
\end{equation}
~\\[-5mm]
or using the notation used for the parameter space in Section \ref{sec:model},
$$
\gamma_n(C) = \theta(u(T)) - \theta(u(0)) ~.
$$

\widetext


\vspace*{-9mm}

\begin{thebibliography}{99}
\vspace*{-18mm}
\narrowtext
\bibitem{zhkm97}
 G.-M. Zhao, M. B. Hunt, H. Keller, and K. A. M\"{u}ller,
 Nature {\bf 385}, 236 (1997).
\bibitem{be93}
 X.-X. Bi and P. C. Eklund,
 Phys. Rev. Lett. {\bf 70}, 2625 (1993).
\bibitem{edsmsyhua94}
 {\it Proceedings of the International
  Workshop on Anharmonic Properties of High-$T_c$ Cuprates}
 (edited by D. Mihailovi\'{c}, G. Ruani, E. Kaldis, and K. A. M\"{u}ller),
 pp. 118-126 (World Scientific, Singapore, 1994).
\bibitem{kim}
 Y. H. Kim, A. J. Heeger, L. Acedo, G. Stucky, and F. Wudl,
 Phys. Rev. B {\bf 36}, 7252 (1987);
 Y. H. Kim, C. M. Foster, A. J. Heeger, S. Cox, and G. Stucky,
 {\it ibid.} {\bf 38}, 6478 (1988).
\bibitem{hshmj97}
 D. Haskel, E. A. Stern, D. G. Hinks, A. W. Mitchell, and J. D. Jorgensen,
 Phys. Rev. B {\bf 56}, R521 (1997).
\bibitem{berry84}
 M. V. Berry, Proc. R. Soc. Lond. A {\bf 392}, 45 (1984).
\bibitem{msw89}
 {\it Geometric Phases in Physics}
 (edited by A. Shapere and F. Wilczek), pp. 160-183
 (World Scientific, Singapore, 1989).
\bibitem{nakahara90}
 M. Nakahara,
 {\it Geometry, Topology, and Physics},
 section 10.6, pp. 364-372 (Adam Hilger, Bristol and New York, 1990).
\bibitem{prz87}
 P. Prelov\v{s}ek, T. M. Rice, and F. C. Zhang,
 J. Phys. C: Solid State Phys. {\bf 20}, L229 (1987).
\bibitem{zs92}
 J. Zhong and H.-B. Sch\"{u}ttler,
 Phys. Rev. Lett. {\bf 69}, 1600 (1992).
\bibitem{rfb93}
 H. R\"{o}der, H. Fehske, and H. B\"{u}ttner,
 Phys. Rev. B {\bf 47}, 12 420 (1993).
\bibitem{frmb93}
 H. Fehske, H. R\"{o}der, A. Mistriotis, and H. B\"{u}ttner,
 J. Phys.: Condens. Matter {\bf 5}, 3565 (1993).
 \bibitem{dagotto94}
 E. Dagotto, Rev. Mod. Phys. {\bf 66}, 763 (1994).
\bibitem{ki85}
  H. Kuratsuji and S. Iida, Prog. Theor. Phys. {\bf 74}, 439 (1985).
\bibitem{berry90}
 M. V. Berry, Physics Today, 34 (December 1990).
\bibitem{aitchison88}
 I. J. R. Aitchison, Physica Scripta T{\bf 23}, 12 (1988).
\bibitem{mt79}
 C. A. Mead and D. G. Truhlar, J. Chem. Phys. {\bf 70}, 2284 (1979).
\bibitem{hl63}
 H. C. Longuet-Higgins, Proc. R. Soc. Lond. A {\bf 344}, 147 (1975);
 G. Herzberg and H. C. Longuet-Higgins,
 Disc. Farad. Soc. {\bf 35}, 77 (1963).
\bibitem{syz95}
 H.-B. Sch\"{u}ttler, K. Yonemitsu, and J. Zhong,
 J. Supercond. {\bf 8}, 555 (1995).
\bibitem{fradkin91}
 E. Fradkin, {\it Field Theories of Condensed Matter Systems},
 section 7.8, pp. 220-221 (Addison-Wesley, 1991).
\widetext
\end{thebibliography}
\end{document}